\documentclass[%
 aip,
 amsmath,amssymb,
preprint,%
]{revtex4-1}

\usepackage{import}
\usepackage{graphicx}
\usepackage{dcolumn}
\usepackage{bm}

\usepackage[utf8]{inputenc}
\usepackage[T1]{fontenc}
\usepackage{mathptmx}
\usepackage{etoolbox}
\usepackage{comment}

\usepackage{lineno}
\usepackage{siunitx}
\usepackage{todonotes}
\usepackage[utf8]{inputenc}
\usepackage[nameinlink]{cleveref}

\newcommand{\mytilde}{\raise.17ex\hbox{$\scriptstyle\mathtt{\sim}$}}

\makeatletter
\def\@email#1#2{%
 \endgroup
 \patchcmd{\titleblock@produce}
  {\frontmatter@RRAPformat}
  {\frontmatter@RRAPformat{\produce@RRAP{*#1\href{mailto:#2}{#2}}}\frontmatter@RRAPformat}
  {}{}
}%
\makeatother
\begin{document}


\title{Controlling the recovery time of the superconducting nanowire single-photon detector with a voltage-controlled cryogenic tunable resistor} 
\author{H. Wang}
 \email{h.wang-19@tudelft.nl}
\affiliation{Department of Imaging Physics, Delft University of Technology, 2628CN Delft.}
\author{N. D. Orlov}%

\affiliation{Department of Imaging Physics, Delft University of Technology, 2628CN Delft.
}%
\author{N. Noordzij}
\affiliation{Single Quantum B.V., 2628 CH Delft, The Netherlands.
}%

\author{T. Descamps}
\affiliation{Department of Applied Physics, Royal Institute of Technology (KTH), SE-106 01 Stockholm, Sweden.}

\author{J. W. N. Los}
\affiliation{Single Quantum B.V., 2628 CH Delft, The Netherlands.}

\author{V. Zwiller}
\affiliation{Department of Applied Physics, Royal Institute of Technology (KTH), SE-106 01 Stockholm, Sweden.}

\author{I. Esmaeil Zadeh}
\altaffiliation[Also at ]{Single Quantum B.V., 2628 CH Delft, The Netherlands.}
\affiliation{Department of Imaging Physics, Delft University of Technology, 2628CN Delft.}

\date{\today}

\begin{abstract}
Superconducting nanowire single-photon detectors (SNSPD), owing to their unique performance, are currently the standard detector in most demanding single-photon experiments. One important metric for any single-photon detector is the deadtime (or recovery time), defined as the minimum temporal separation between consecutive detection events. In SNSPDs, the recovery time is more subtle, as the detection efficiency does not abruptly drop to zero when the temporal separation between detection events gets smaller, instead, it increases gradually as the SNSPD current recovers.  SNSPD's recovery time is dominated by its kinetic inductance, the readout impedance, and the degree of saturation of internal efficiency. Decreasing the kinetic inductance or increasing the readout impedance can accelerate the recovery process. Significant reduction of the SNSPD recovery time, by, for example, adding a series resistor in the readout circuitry, is possible but can lead to detector latching which hinders further detector operation or enforces underbiasing and hence a reduction in detection efficiency. Previous research has demonstrated passive resistive networks for the reduction of recovery time that rely on trial and error to find the appropriate resistance values. Here we show, using a novel, cryogenically compatible, and tunable resistor technology, one can find the optimized impedance values, delivering fast SNSPD recovery time, while maintaining maximum internal detection efficiency. Here we show an increase of more than 2 folds in both maximum achievable detection rates and the achievable detection efficiency at high photon fluxes, demonstrating detection rates as high as 120 Mcps with no loss of internal detection efficiency.  
\end{abstract}

\maketitle

\begin{figure*}[htbp]
    \centering
    \includegraphics[width =0.75\textwidth]{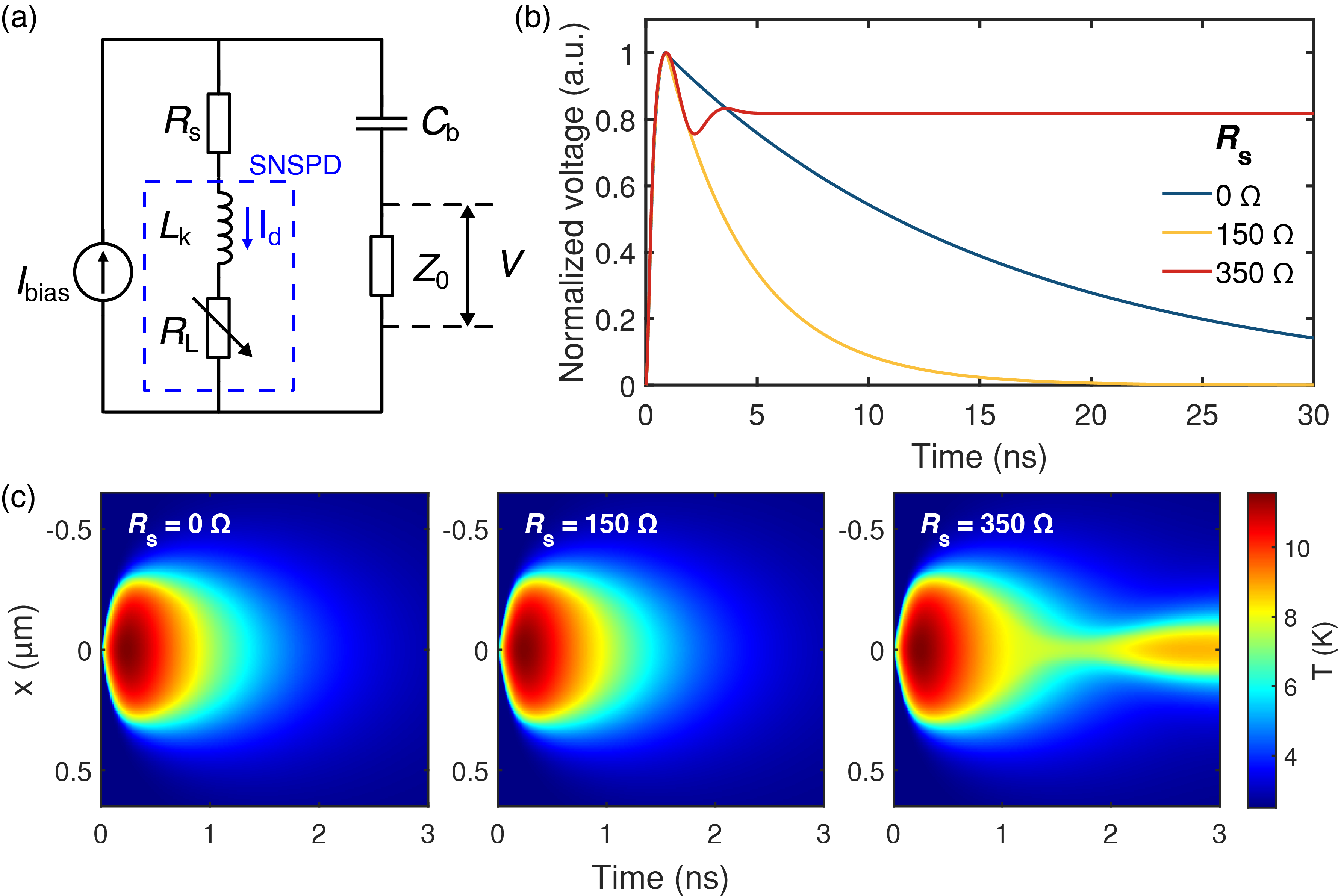}
    \caption{(a) Schematic representation of electrical equivalent model of a SNSPD with a resistor $R_\mathrm{s}$ in series. (b) Simulated output voltage profile with $R_\mathrm{s}=$ \qty{0}{\Omega}, \qty{150}{\Omega} and \qty{350}{\Omega}, respectively. (c) The evolution of the temperature profile of the superconducting nanowire with $R_\mathrm{s}=$ \qty{0}{\Omega}, \qty{150}{\Omega} and \qty{350}{\Omega}, respectively. Details of the simulation parameters can be found in supplementary material section II. }
    \label{fig:sim_recovery_time}
\end{figure*}
\begin{figure}[htbp]
    \centering
    \includegraphics[width =0.45\textwidth]{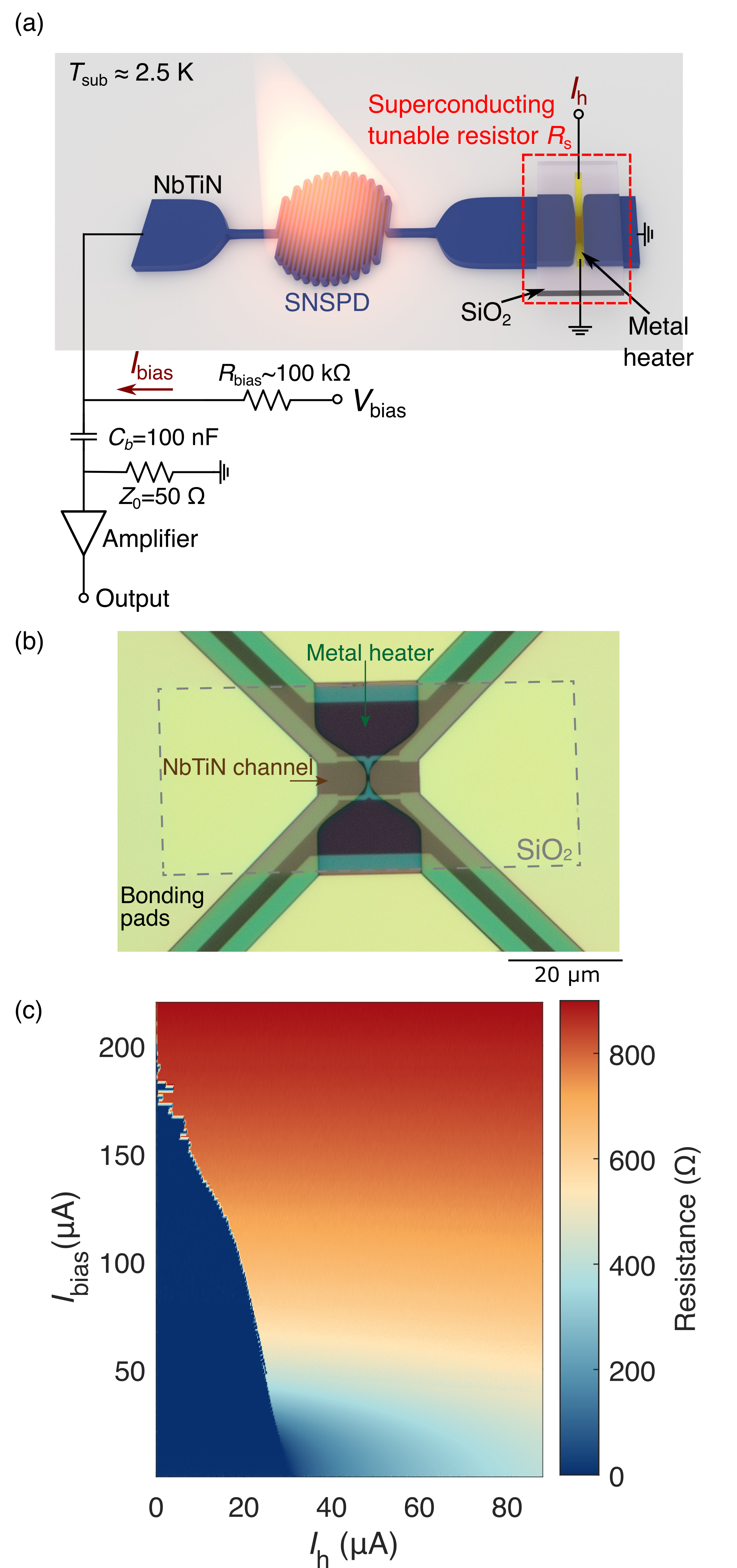}
    \caption{(a) Illustration of a cryogenic tunable resistor $R_\mathrm{s}$ in series with a SNSPD and its electronic readout circuit. (b) Representative optical microscopy image of a cryogenic tunable resistor.  (c) Characterization of the channel resistance in a cryogenic tunable resistor as a function of the channel bias current and the heater current. The channel has a width of \qty{1}{\micro\meter} and the heater has a width of \qty{100}{\nano\meter}.}

    \label{fig:characterization}
\end{figure}

Superconducting nanowire single-photon detectors (SNSPDs), composed of superconducting nanowires biased close to their critical currents, have become a cutting-edge technology in single-photon detection.\cite{Gol2001,Lau2023,Venza2025} SNSPDs have gained numerous research interests, attributed to their remarkable single photon sensitivity, broad spectral range\cite{Pan2022,Colangelo2022}, high detection efficiency (>\qty{95}{\percent}\cite{Chang2021,Reddy2020,Hu2020}), and a low time jitter (tens of picoseconds\cite{Korzh2020,Taylor2022,Chang2019}). There is an increasing demand for SNSPDs in various applications such as biomedical imaging\cite{Xia2021}, optical communication\cite{Bellei2016}, and quantum key distribution\cite{Takemoto2015,Fadri2023,Beutel2021}. \\
\indent When a photon is absorbed by the SNSPD, a resistive section is formed in the nanowire, which will initially expand due to Joule heating (current passing through a resistor). Subsequently, in an appropriately functioning SNSPD, the current is diverted to the readout, and this resistive section disappears due to the reduction of Joule heating and continuous cooling of the sample. This dynamics in the current redistribution between the SNSPD and the readout electronics leads to an electrical pulse at the output. The resulting output pulse usually rises steeply (sub-nanosecond) and decays slowly (nanoseconds), which is correlated with the inductive time constant $L_\mathrm{k}/R$ ($L_\mathrm{k}$ is the kinetic inductance of the SNSPD, and $R$ is the resistance in the circuit). Thus, the falling edge of the output influences the nanowire current recovery time, which is linked to its internal detection efficiency as a result. As the efficiency does not drop and recover abruptly in SNSPDs, the recovery time $\tau_\mathrm{rec}$ can be defined as the time interval required to register a second photon subsequent to an earlier detection event with $\ge$ \qty{50}{\%} internal detection efficiency. This parameter is one of the essential performance metrics for applications that demand higher photon count rates, such as optical communication, or high-throughput optical quantum computing.\\
\indent SNSPDs have a shorter recovery time when their kinetic inductances are lower. However, this also results in the reduction of the active detection area, which is disadvantageous for optical absorption and hence system detection efficiency. Other approaches to raise detection rates, such as utilizing Quasi-Constant-Voltage (QCV) bias\cite{Liu2012}, employing the resistive attenuator readout circuit\cite{Zhao2014}, and implementing a DC-coupled readout scheme\cite{Kerman2013}, were demonstrated, but these non-cryogenic solutions have limited performance enhancements as there is a considerable delay (nano-seconds) between the SNSPD and the control circuit. Another more effective way is to introduce a cryogenic serial resistor to SNSPD's readout circuit to reduce the recovery time by increasing $R$.\cite{Zhao2014,Lv2017} The value of the serial resistance needs to be carefully optimized to minimize recovery time while avoiding the formation of a stable hotspot in the superconducting nanowire, which is referred to as the latching phenomenon\cite{Yang2007}. Based on the analysis of electrothermal effects in SNSPDs, the optimized value for this serial resistance can be influenced by various factors such as the heat transfer coefficient between the superconducting nanowire and the substrate, the geometry of the meandering nanowire, and the material properties of the superconducting layer (sheet resistance, critical current density, biasing current, etc.).\cite{Kerman2009} Therefore, the resistance value is often refined experimentally through a series of trials and errors, which is time-consuming and does not necessarily guarantee optimal performance.\\
\indent In this work, we demonstrate a cryogenic tunable resistor to tune the serial resistance to the optimal value. The tunable resistor consists of a superconducting channel and a metal heater insulated by a $\mathrm{SiO}_2$ layer. The resistance of the superconducting channel is controlled by the localized heat transfer from the heater and the self-Joule heating of the superconducting channel. This adjustable electrical component operating in cryogenic environments constitutes an efficient strategy to enhance the SNSPD performance at high photon flux rates. We show that this on-chip device is fully compatible with SNSPDs, enabling real-time optimization of the electrical recovery time while maintaining high internal detection efficiency.\\
\indent Fig. \ref{fig:sim_recovery_time} represents a one-dimensional electrothermal simulation as described in \citet{Yang2007}, which illustrates the basic principle to decrease the recovery time by connecting a small resistor in series with a SNSPD (More information about the simulation can be found in supplementary material section II). The results were simulated at \qty{2.5}{K}, which is the base temperature of our cryostat. An equivalent electrical diagram of a SNSPD in series with a resistor $R_\mathrm{s}$ is depicted in Fig. \ref{fig:sim_recovery_time}(a). The SNSPD is biased with a current source $I_\mathrm{bias}$ and is in parallel with the readout electronics, which is a capacitor $C_\mathrm{b}$ and a load resistor $Z_0=$ \qty{50}{\ohm}, representing the input impedance of an RF amplifier. The simple circuit here, ignores the time delay (corresponding to the transmission lines and the coaxial cables) between the detector and the amplifier. This provides an acceptable approximation as long as the recovery time is longer than the mentioned time delay (normally on the order of a few nanoseconds). \\
\indent Once a single photon triggers a resistive section across the nanowire, this section quickly expands due to Joule heating and the bias current through the detector $I_\mathrm{d}$ is redirected to the load resistor, resulting in the rising edge at the output. The characteristic time of the rising edge is $\tau_1(t)\approx L_\mathrm{k}/(Z_0+R_\mathrm{s}+R_\mathrm{L}(t))$, where $L_\mathrm{k}=$ \qty{750}{\nH} and $R_\mathrm{L}(t)$ represent the kinetic inductance of the detector and the real-time total hot-spot resistance in our simulation. After the resistive spot disappears via the heat dissipation to the substrate, the current flows back to the SNSPD with the time constant $\tau_2\approx L_\mathrm{k}/(Z_0+R_\mathrm{s})$, corresponding to the falling edge of the output signal.
Since $R_\mathrm{L}(t)$ (typically a few k$\Omega$) significantly exceeds $Z_0$ and $R_\mathrm{s}$, the pulse duration $\tau_\mathrm{out}$ defined at $1/e$ of the pulse height is dominated by $\tau_2$.\\
\indent The simulated output signals with three different $R_\mathrm{s}$ values are shown in Fig. \ref{fig:sim_recovery_time}(b). The pulse decay time (estimated via exponential fitting), reduces from \qty{14.99}{\ns} to \qty{3.75}{\ns} by increasing $R_\mathrm{s}$ from \qty{0}{\ohm} to \qty{150}{\ohm} without changing the nanowire structure, which is consistent with the analytical expression of $\tau_\mathrm{2}$. However, for a larger value of $R_\mathrm{s}=$ \qty{350}{\Omega}, the detector latches to a non-superconducting state, as represented by the temperature profile in Fig. \ref{fig:sim_recovery_time}(c). In the latter case, the current returns to the SNSPD so rapidly that the resistive spot is sustained by reaching an equilibrium between heat generation via Joule heating in the nanowire and heat dissipation to the substrate. The speed of current redistribution depends on the kinetic inductance of the SNSPD, and thermal interactions are related to the heat transfer coefficient between the nanowire and the substrate. Unfortunately, measuring these properties accurately on different platforms is not straightforward and, therefore, it is non-trivial to predict the maximum $R_\mathrm{s}$ which optimizes the current recovery while avoiding the latching or loss of internal efficiency. Using fixed resistors, the experimental testing and validation process can take a considerable amount of effort and time. Hence, here we propose and demonstrate a cryogenic tunable resistor, as depicted in Fig. \ref{fig:characterization}, which makes it convenient to optimize the SNSPD performance with a serial tunable resistor.  \\
\indent Fig. \ref{fig:characterization}(a) shows the schematic of a tunable resistor. A metal strip (a \qty{80}{\nm}-thick Ti layer with a \qty{5}{nm}-thick Au layer) is fabricated over a superconducting channel (NbTiN, with a thickness of 8-10 nm) insulated by a thin SiO$_2$ layer (40-50 nm in thickness). The metal strip acts as a micro-heater once connected to an electric power supply, producing Joule heat and raising the local temperature. When the local temperature exceeds the critical temperature, a resistive strip is formed in the channel, which introduces a tunable resistance in the circuit. The channel is narrowed down as it approaches the center of the metal heater to constrict the expansion of the resistive area and prevent the channel from latching. A typical optical micrograph of a cryogenic tunable resistor is shown in Fig. \ref{fig:characterization}(b). \\
\indent The resistance of the channel is determined by both the bias current through the superconducting channel $I_\mathrm{ch}$, and the heater current $I_\mathrm{h}$. In Fig. \ref{fig:characterization}(c), we characterize the dependence of channel resistance $R$ on the currents $I_\mathrm{ch}$ and $I_\mathrm{h}$ in the device (DC analysis). The minimum width of the NbTiN channel is \qty{1}{\micro\meter} and the metal heater has a width of \qty{0.1}{\micro\meter}. As the bias current through the superconducting channel increases, the resistance grows faster with the heater current. This behavior is due to the generation of Joule heat in the NbTiN channel $I_\mathrm{ch}^2R$, which becomes significant at higher bias currents. To ensure reliable tunability in the desired resistance range (0 - 300 $\mathrm{\Omega}$), the tunable resistor is designed to operate at bias currents much lower than the critical current of the channel, typically around \qty{10}{\%} depending on the heater and the channel geometry as well as the resistivity of the superconducting material. 
\begin{figure*}[htbp]
    \centering
    \includegraphics[width =0.7\textwidth]{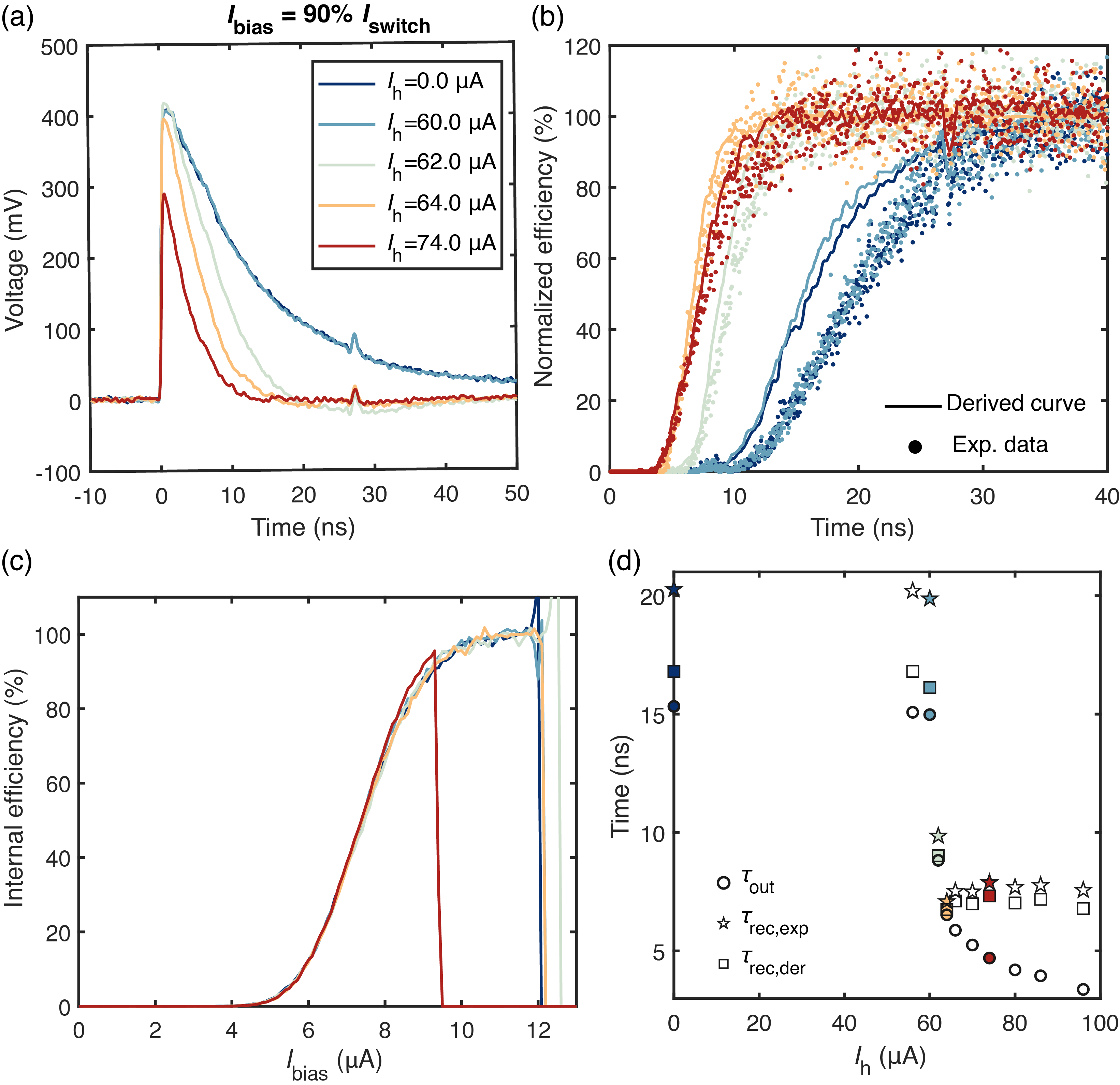}
    \caption{(a) Amplified output waveforms of the SNSPD (curves are averaged over 100 sweeps). (b) Recovery of the detection probability after one detection event with various $I_\mathrm{h}$ for the wavelength of \qty{1548}{\nano\meter}. The filled dots represent the experimental data, and the solid ones are the predicted efficiency recovery curves derived from internal detection efficiency curves and output pulses. (c) Normalized internal detection efficiency curves with various heater currents $I_\mathrm{h}$, obtained for the wavelength of \qty{1548}{\nano\meter}. (d) Comparison of the output pulse recovery times width $\tau_\mathrm{out}$ (measured at $1/e$ of the pulse amplitude), the experimental recovery time $\tau_\mathrm{rec,exp}$, and the derived recovery time $\tau_\mathrm{rec,der}$. $\tau_\mathrm{rec,exp}$ and $\tau_\mathrm{rec,der}$ are obtained by fitting the data with the sigmoid function $y=A(1+\mathrm{erf}(S(x-x_\mathrm{0})))$.}
    \label{fig:exp-recovery-time_new1548}
\end{figure*}\\
\indent As depicted in Fig. \ref{fig:characterization}(a), a SNSPD is connected in series with a cryogenic tunable resistor and biased and read out using a conventional bias tee and RF amplifiers. We characterize the sample in flood illumination configuration with light injected from a fiber several centimeters above the sample. The output pulse is amplified and recorded using an oscilloscope. Both the theoretical analysis and the simulation results indicate that a higher $I_\mathrm{h}$ results in a shorter output pulse duration $\tau_\mathrm{out}$, allowing two consecutive photons to, in principle, be detected with a shorter time delay. However, another crucial characteristic to consider is the detection efficiency of the second event, which itself depends on not only the current recovery speed but also the degree of saturation of internal detection efficiency (IDE) (i.e., the relative length of the plateau in the detection efficiency curve to the switching current). Since the detection efficiency of SNSPD increases with the effective bias current, the probability of detecting a second photon grows as the current through the SNSPD recovers. Thus, we experimentally measured the recovery of the detection efficiency as a function of the time delay after the first detection event using continuous light sources at different heater currents (i.e., different serial resistances). The time when the efficiency recovers to \qty{50}{\%} is defined as the recovery time $\tau_\mathrm{rec}$.
\\
\indent Fig. \ref{fig:exp-recovery-time_new1548} shows the experimental results of a SNSPD composed of \qty{70}{\nm}-wide nanowires in series with a cryogenic tunable resistor with a \qty{1}{\um}-wide channel and a \qty{0.3}{\um}-wide heater. The SNSPD was biased at around \qty{90}{\%} of the switching current and measured at a wavelength of \qty{1548}{\nano\meter}. Fig. \ref{fig:exp-recovery-time_new1548}(a) shows the recorded output pulses for various heater currents $I_\mathrm{h}$. The experimental pulse widths match the trend in the simulation results. As the serial resistance increases by setting a higher heater current, the falling time constant of the output declines from \SI[separate-uncertainty = true]{15.3}{\nano\second} at $I_\mathrm{h}=$ \qty{0}{\micro\ampere} to \SI[separate-uncertainty = true]{3.4}{\nano\second} at $I_\mathrm{h}=$ \qty{74}{\micro\ampere}. \\
\indent However, the efficiency recovery time does not monotonically decrease as the serial resistance increases. We experimentally measured the efficiency recovery curve (filled dots) as shown in Fig. \ref{fig:exp-recovery-time_new1548}(b) (More information about the measuring approach can be found in supplementary material section III). Fig. \ref{fig:exp-recovery-time_new1548}(d) shows the comparison between the pulse durations $\tau_\mathrm{out}$ measured from the averaged waveforms in Fig. \ref{fig:exp-recovery-time_new1548}(a), and the experimental recovery time $\tau_\mathrm{rec,exp}$ fitted using the experimental efficiency recovery curve in Fig. \ref{fig:exp-recovery-time_new1548}(b). The experimental recovery time $\tau_\mathrm{rec,exp}$ is suppressed from \qty{20.3}{\nano\second} at $I_\mathrm{h}=$ \qty{0}{\micro\ampere} to \qty{7.1}{\nano\second} at $I_\mathrm{h}=$ \qty{64}{\micro\ampere}, which is consistent with the declining trend of the output pulse duration. However, the recovery time increases to \qty{7.9}{\nano\second} at $I_\mathrm{h}=$ \qty{74}{\micro\ampere} although $\tau_\mathrm{out}$ becomes shorter. Therefore, there is an optimal heater current $I_\mathrm{h}$ at \qty{64}{\uA}, that is, an optimal serial resistance $R_\mathrm{s}$, where the efficiency recovery time can be minimized by reducing the pulse duration.\\
\indent This results from the degradation of the internal detection efficiency of the detector. Fig. \ref{fig:exp-recovery-time_new1548}(c) shows the internal detection efficiency (IDE) curves as a function of the bias current, which are normalized to the saturated efficiency estimated by fitting to the sigmoid function ($y=A(1+\mathrm{erf}(S(x-x_\mathrm{0})))$)\cite{Autebert2020}. Since $I_\mathrm{h}=$ \qty{74}{\micro\ampere} leads to a lower and unsaturated IDE curve, a higher $I_\mathrm{d}/I_\mathrm{switch}$ ratio, which corresponds to a longer delay between two events, is required to achieve the same detection probability. This effect counteracts the influence of the shorter output pulse duration, leading to a worse efficiency recovery time at $I_\mathrm{h}=$ \qty{74}{\micro\ampere}.   \\
\indent The change in the IDE curves is attributed to the decrease of the switching current $I_\mathrm{switch}=\mathrm{min}\{I_\mathrm{c}, I_\mathrm{latch}\}$, where $I_\mathrm{c}$ is the critical current determined by the geometry of the nanowire and the critical current density of the superconducting film, and $I_\mathrm{latch}$ is the latching current at which the resistive spot in the superconducting nanowire is retained. As the output pulse width $\tau_\mathrm{out}$ decreases, $I_\mathrm{latch}$ declines and eventually drops below $I_\mathrm{c}$.\cite{Kerman2013,Craiciu2023} This results in the reduction of $I_\mathrm{switch}$, restricting the bias current from approaching the critical current. Consequently, the saturation plateau of the internal detection efficiency diminishes as $R_\mathrm{s}$ exceeds an optimal value. The switching current increases slightly at $I_\mathrm{h}=$ \qty{62}{\micro\ampere}, which may result from impedance matching between the tunable resistor and the readout load resistor.\\
\indent Furthermore, we investigate the recovery of the instantaneous detection efficiency derived from the output pulses and the IDE curve in Fig. \ref{fig:exp-recovery-time_new1548}(a) and (c).\cite{Oshiro2025} The instantaneous effective bias current can be estimated using $I_\mathrm{d}(t)=I_\mathrm{bias}-V_\mathrm{out}(t)/G/Z_0$, where $V_\mathrm{out}(t)$ is the output pulse and $G=10^{2.9}$ is the measured gain of our amplifier. Combined with the IDE curve $\eta=f(I_\mathrm{d})$, the instantaneous detection efficiency can be derived as $\eta(t)=f(I_\mathrm{d}(t))$, which is shown as solid lines in Fig. \ref{fig:exp-recovery-time_new1548}(b). The derived recovery time $\tau_\mathrm{rec,der}$ is determined using sigmoidal fitting and plotted in Fig. \ref{fig:exp-recovery-time_new1548}(d). Discrepancies between $\tau_\mathrm{rec,der}$ and $\tau_\mathrm{rec,exp}$ are probably caused by inaccurate estimations of effective bias currents, and the triggering threshold for the second detection event (hysteresis settings of counter's Schmitt trigger), which omits some events because of their lower amplitude or background noise (see more details in supplementary material section III). Nevertheless, $\tau_\mathrm{rec,der}$ and $\tau_\mathrm{rec,exp}$ show similar tendencies with respect to $I_\mathrm{h}$, and exhibit good agreement under conditions of rapid efficiency recovery. Therefore, analyzing $\tau_\mathrm{rec,der}$ can be considered as a time-efficient approach to optimize the serial tunable resistor.\\
\begin{figure*}[htbp]
    \centering
    \includegraphics[width =0.9\textwidth]{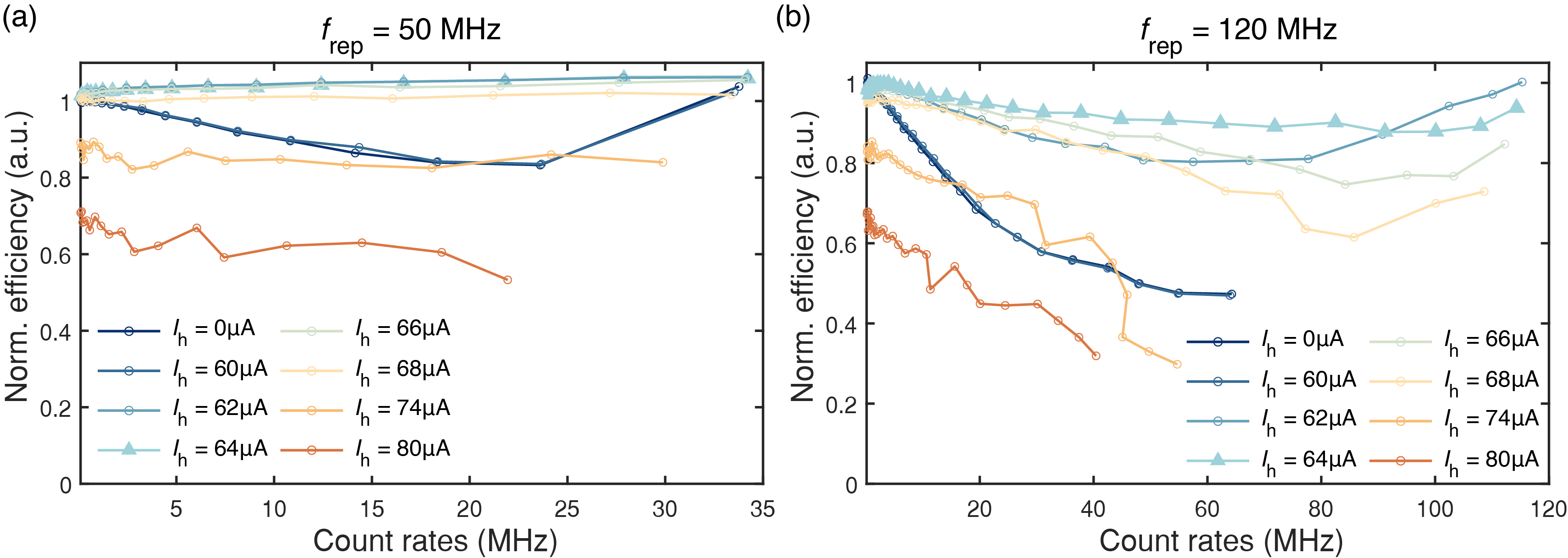}
    \caption{Normalized detection efficiency as a function of detection count rates at various heater currents $I_\mathrm{h}$ with a repetition rate of (a) $f_\mathrm{rep}=$ \qty{50}{\MHz} and (b) $f_\mathrm{rep}=$ \qty{120}{\MHz}. For each laser repetition rate, the best detector performance can be achieved by tuning the heater current. Measurements at $I_\mathrm{h}=$ \qty{64}{\uA}, which consistently delivers near-optimal or optimal performances across a broad range of count rates, are highlighted with triangles for a better display.}
    \label{fig:HCR1}
\end{figure*}
\indent In order to further investigate the effect of recovery time reduction, we characterize the SNSPD detection efficiency at high count rates illuminated with a pulsed light source at \qty{1548}{\nm} at different heater currents. The detection efficiency, for each input photon flux, is calculated using the count rates at \qty{95}{\%} of the switching current. Examples of the measurement curves (count rate versus bias current at different input photon fluxes) are presented in supplementary material section IV. The analyzed results of are shown in Fig. \ref{fig:HCR1} for the laser repetition rates of \qty{50}{\MHz} and \qty{120}{\MHz}. Note that the data in Fig. \ref{fig:HCR1} is normalized with respect to the efficiency at for the lowest input photon flux (lowest count rate) when $I_\mathrm{h}=$ \qty{0}{\uA} (more measurements at different repetition rates can be found in supplementary material).\\
\indent According to the measurement in Fig. \ref{fig:exp-recovery-time_new1548}(d), the recovery time of the SNSPD is longer or comparable to $1/f_\mathrm{rep}$ at lower heater currents than \qty{60}{\uA}. Thus, the detection efficiencies for these lower heater currents decrease as the count rate increases (due to the insufficient current recovery through the SNSPD). When the count rates increase beyond half of $f_\mathrm{rep}$, the efficiency exhibits an upward tendency, which is attributed to the increase of SNSPD current via parasitic discharging currents of the readout capacitor.\cite{Ferrari2019} The parasitic currents flowing back through the SNSPD enhance the effective bias currents and thus the detection efficiency.\\ 
\indent As the heater current increases, the best efficiency performance over a broad range of count rates is achieved at $I_\mathrm{h}=$ \qty{64}{\uA}, which is in accordance with the recovery time measurement. When the heater current exceeds \qty{64}{\uA}, the efficiency declines steadily due to the unsaturated IDE (see supplementary material Fig. S4). At $f_\mathrm{rep}=$ \qty{50}{\MHz} and $I_\mathrm{h}=$ \qty{64}{\uA}, the detection efficiency is maintained around the highest level at various count rates, since the minimal time delay between two pulses, which is \qty{20}{\ns}, is significantly longer than the recovery time $\tau_\mathrm{rec,exp}=$ \qty{7.1}{ns}, resulting in complete recovery of the effective bias current for all detection events. As the minimum separation between incident photons for the case of $f_\mathrm{rep}=$ \qty{120}{\MHz} approaches the recovery time (at $I_\mathrm{h}=$ \qty{64}{\uA}), and the efficiency recovery is not fully completed, a slight decline in efficiency can be observed at higher count rates. 

\indent 
In conclusion, we demonstrated a cryogenic tunable resistor controlled by the localized micro-heater. The device is intrinsically cryogenic-compatible and suitable for integration with other superconducting devices. We showed that, by controlling the tunable resistor in series with a SNSPD, the efficiency recovery time of the detector can be tuned to enhance detection efficiency at high count rates. Compared with trial-and-error-based fixed resistive networks, the cryogenic tunable resistor significantly speeds up the optimization process and can be adapted to a variety of SNSPD geometries or materials. Our solution provides a convenient method to tune SNSPD performance at higher photon fluxes. In addition, the tunable resistor presented here has the potential to be integrated with other cryogenic technologies. 
\\

See the supplementary material for details on sample fabrication, electro-thermal simulation of SNSPDs, the approach to measure the efficiency recovery, and more results of high-count-rate measurements.
\begin{acknowledgments}
I. E. Z. and H. W. acknowledge the funding from the FREE project (P19-13) of the TTW-Perspectief research program partially financed by the Dutch Research Council (NWO); I. E. Z. acknowledges funding from the European Union’s Horizon Europe research and innovation programme under grant agreement No. 101098717 (RESPITE project) and No.101099291 (fastMOT project). 
\end{acknowledgments}
\section*{Author Declarations}
\subsection*{Conflict of Interest}
The authors have no conflicts to disclose.
\subsection*{Author Contributions}
\textbf{Hui Wang:} Conceptualization (equal); Data curation (equal); Formal analysis (equal); Investigation (lead); Methodology (equal); Resources (equal); Software (equal); Supervision (equal); Validation (equal); Visualization (equal); Writing – original draft (lead); Writing – review \& editing (equal). \textbf{Nikita Dmitrievič Orlov:} Data curation (equal); Formal analysis (equal); Investigation (equal); Writing – original draft (equal); Writing – review \& editing (equal). \textbf{Niels Noordzij:} Investigation (equal); Methodology (equal); Resources (equal); Writing – original draft (equal); Writing – review \& editing (equal). \textbf{Thomas Descamps:} Investigation (equal); Resources (equal); Writing – original draft (equal); Writing – review \& editing (equal). \textbf{J. W. Niels Los:} Conceptualization (equal); Supervision (equal); Validation (equal); Writing – original draft (equal); Writing – review \& editing (equal). \textbf{Val Zwiller:} Supervision (equal); Writing – original draft (equal); Writing – review \& editing (equal). \textbf{Iman Esmaeil Zadeh:} Conceptualization (equal); Funding acquisition (equal); Investigation (equal); Methodology (equal); Project administration (equal); Supervision (equal); Validation (equal); Visualization (equal); Writing – original draft (equal); Writing – review \& editing (equal).
\section*{Data Availability}
The data that support the findings of this study are available on request from the authors.

\bibliography{bibl}

\end{document}


\preprint{AIP/123-QED}

\title{Supplementary material: Controlling the recovery time of the superconducting nanowire single-photon detector with a low-power and voltage-controlled cryogenic tunable resistor}
\author{H. Wang}
 \email{h.wang-19@tudelft.nl}
\affiliation{Department of Imaging Physics, Delft University of Technology, 2628CN Delft.}
\author{N. D. Orlov}%

\affiliation{Department of Imaging Physics, Delft University of Technology, 2628CN Delft.
}%
\author{N. Noordzij}
\affiliation{Single Quantum B.V., 2628 CH Delft, The Netherlands.
}%
\author{T. Descamps}
\affiliation{Department of Applied Physics, Royal Institute of Technology (KTH), SE-106 01 Stockholm, Sweden.}

\author{J. W. N. Los}
\affiliation{Single Quantum B.V., 2628 CH Delft, The Netherlands.}

\author{V. Zwiller}
\affiliation{Department of Applied Physics, Royal Institute of Technology (KTH), SE-106 01 Stockholm, Sweden.}

\author{I. Esmaeil Zadeh}
\altaffiliation[Also at ]{Single Quantum B.V., 2628 CH Delft, The Netherlands.}
\affiliation{Department of Imaging Physics, Delft University of Technology, 2628CN Delft.}
{
\let\clearpage\relax
\maketitle
}
\section{Sample fabrication}
The deposition of NbTiN films is carried out on thermally oxidized silicon wafers in magnetron cosputtering in Ar and $\mathrm{N}_2$ atmosphere. The film thickness is around \qty{10}{\nm}, corresponding to the critical temperature at around 8$\sim$10 K (see \citet{Tcmeas}). Next, the gold contact pads and markers are patterned using electron-beam lithography (EBL, 100 kV acceleration) with a double-layer PMMA resist. After development in a 1:3 MIBK:IPA solution, a 5-nm Cr sticking layer and a 55-nm Au layer are deposited via electron-beam metal evaporation. Lift-off is then carried out in acetone at \qty{45}{\degreeCelsius}. Afterwards, NbTiN patterns are created with EBL using the positive resist ARP6200.4 and developed in pently acetate. The NbTiN structures are later etched using reactive ion etching (RIE) with an $\mathrm{SF}_6/\mathrm{O}_2$ plasma. $\mathrm{SiO}_2$ insulating structures are then patterned using a positive resist ARP6200.9. After deposition of a 30$\sim$40 nm-thick $\mathrm{SiO}_2$ layer in inductively coupled plasma chemical vapor deposition (ICPCVD) at \qty{150}{\degreeCelsius}, the structures are lifted off in PRS-3000. Finally, the metal heaters are fabricated through another lift-off process using double-layer PMMA resist. They are composed of a 80 nm-thick Ti layer and a 5 nm-thick Au layer deposited using electron-beam metal evaporation. 
\begin{figure}[htbp]
    \centering
    \includegraphics[width =1\textwidth]{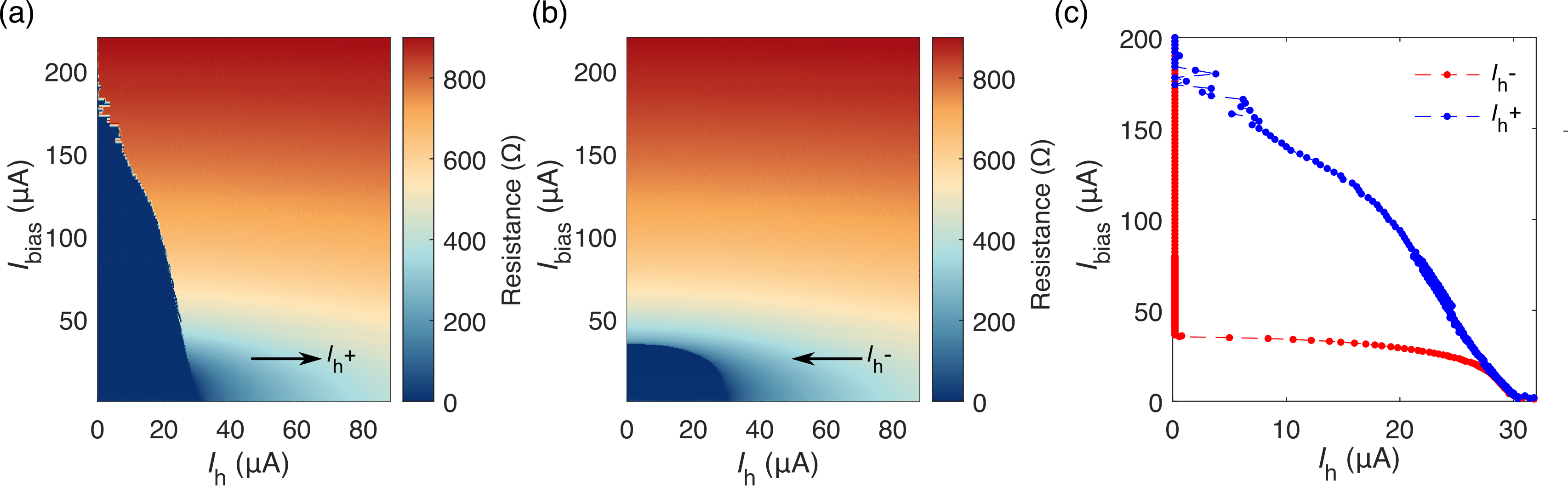}
    \caption{(a) and (b) Hysteresis $I_\mathrm{h}$-$R$ measurement of a cryogenic tunable resistor with $I_\mathrm{h}$ sweeping in the (a) increasing, and (b) decreasing direction. The NbTiN channel is \qty{1}{\um} wide and the heater is \qty{100}{\nm} wide. (c) The switching heating current ($I_\mathrm{h}+$) and the retrap current ($I_\mathrm{h}-$) of the NbTiN channel obtained from the results in (a) and (b). }
    \label{fig:hysteresis}
\end{figure}

\section{Electro-thermal simulation of SNSPDs}
\label{si:sim}
We conduct a 1D electro-thermal simulation of a SNSPD with a width of $w=$ \qty{70}{\nm} and a length of $l=$ \qty{525}{\um} using the coupled electro-thermal differential equation described by \textcite{Yang2007}. The equivalent electrical circuit of the simulation is shown in Fig. 1(a). The coupled electrothermal equations are expressed as\cite{Yang2007}
\begin{empheq}[left=\empheqlbrace]{align} 
        &J^2\cdot\rho+\kappa\frac{\partial^2T}{\partial x^2}-\frac{\alpha}{d}(T-T_\mathrm{sub})=\frac{\partial cT}{\partial t},
    \label{eq:thermal}\\
       &C_\mathrm{b}\left( \frac{d^2L_\mathrm{k}I}{dt^2}+\frac{d(IR_\mathrm{L})}{dt}+(Z_\mathrm{0}+R_\mathrm{S})\frac{dI}{dt}\right)=I_\mathrm{bias}-I.
    \label{eq:electrical}
    \end{empheq}
In the thermal differential equation (Equation \ref{eq:thermal}), $J$ is the current density, $\rho$ is the resistivity, $\kappa$ is the thermal conductivity of the superconducting film, $\alpha$ is the thermal boundary conductivity between the superconducting film and the substrate, $d$ is the film thickness, $T_\mathrm{sub}$ is the base temperature, and $c$ is the heat capacity of the superconducting material.\\
In the electrical differential equation (Equation \ref{eq:electrical}), $C_\mathrm{b}=$ \qty{100}{\nF} is the capacitance, $L_\mathrm{k}$ is the kinetic inductance of the SNSPD, $R_\mathrm{L}$ is the resistance of the SNSPD, $Z_\mathrm{0}=$ \qty{50}{\ohm} is the load resistor in readout electronics, and $R_\mathrm{S}$ is the serial resistance. Equation \ref{eq:thermal} and \ref{eq:electrical} are discretized using the Crank-Nicolson method and then solved as a tridiagonal-matrix problem in MATLAB.  \\
The film properties used in the simulation are derived from either literature or experimental data. The film thickness is approximated as $d=$\qty{10}{\nm} and the critical temperature is assumed as $T_\mathrm{c}=$\qty{10}{\kelvin}. The base temperature is set to $T_\mathrm{sub}=$ \qty{2.5}{\kelvin}. The sheet resistance and the sheet inductance of the NbTiN film are estimated as $R_\mathrm{d}=\rho/d=$\num{300} $\Omega/\square$ and $L_\mathrm{k,d}=$\num{100} $\mathrm{pH}/\square$. Therefore, the kinetic inductance of the SNSPD is $L_\mathrm{k}=L_\mathrm{k,d}l/w=$ \qty{750}{\nH}. The critical current of the SNSPD follows
\begin{equation}
    I_\mathrm{c}(T)=I_\mathrm{c}(T=0)\left(1-\left(\frac{T}{T_\mathrm{c}}\right)^2\right)^2,
\end{equation}
where $I_\mathrm{c}(T=0)$ is estimated as \qty{13.13}{\uA} (which is derived from the experimental critical current $I_\mathrm{c}(T=T_\mathrm{sub})=$ \qty{12.1}{\uA}). The SNSPD is biased with $I_\mathrm{bias}=0.9I_\mathrm{c}(T=T_\mathrm{sub})$ \qty{10.89}{\uA} in the simulation.\\
The thermal conductivity $\kappa$ relies on both the state and the temperature, which can be described as
\begin{equation}
    \kappa(T)=
    \begin{cases}
    LT/\rho,& \text{in normal state,}\\
    \frac{T}{T_\mathrm{c}}\kappa(T_\mathrm{c})=\frac{LT^2}{T_\mathrm{c}\rho},& \text{in superconducting state},
    \end{cases}
\end{equation}
where $L=$ \qty{2.45e-8}{\watt\ohm\per\square\kelvin} is the Lorenz number.\\
The thermal boundary conductivity $\alpha$ between the substrate and the NbTiN film depends on the temperature, i.e. $\alpha=BT^3$, according to the literature.\cite{Yang2007} When a hot spot can be maintained in the nanostructure, Equation \ref{eq:thermal} can be rewritten as\cite{DaneSelfheat}
\begin{equation}
    \alpha(T=T_\mathrm{c}^\ast)=\frac{J_\mathrm{retrap}^2d\rho}{T_\mathrm{c}^\ast-T_\mathrm{sub}}=\frac{I_\mathrm{retrap}^2R_\mathrm{d}}{w^2(T_\mathrm{c}^\ast-T_\mathrm{sub})},
    \label{eq:alpha}
\end{equation}
where $T_\mathrm{c}^\ast$ can be derived from the switching current(i.e., the bias current where the superconducting state is about to break) and the retrapping current (i.e., the bias current where the device cannot recover from the resistive state to the superconducting state) 
\begin{equation}
\begin{cases}
    I_\mathrm{retrap}=I_\mathrm{c}(T=0)\left(1-\left(\frac{T_\mathrm{c}^\ast}{T_\mathrm{c}}\right)^2\right)^2,&\\
    I_\mathrm{switch}=I_\mathrm{c}(T=0)\left(1-\left(\frac{T_\mathrm{sub}}{T_\mathrm{c}}\right)^2\right)^2.&
\end{cases}   
\label{eq:alpha-Tc}
\end{equation}
We measured the hysteresis of a cryogenic tunable resistor (with a 1-$\upmu$m wide superconducting channel and a 100-nm wide metal heater) to estimate the parameter $B$ using Equation \ref{eq:alpha} and \ref{eq:alpha-Tc}. As shown in Fig. \ref{fig:hysteresis} (a) and (b), the heater current was swept in ascending or descending order while maintaining the same bias current through the superconducting channel. When there is no heater current, the device can be switched to a non-superconducting state at $I_\mathrm{switch}\approx$ \qty{192}{\micro\ampere}, and be trapped in a resistive state at $I_\mathrm{retrap}\approx$ \qty{36}{\micro\ampere}. By solving Equation \ref{eq:alpha} and Equation \ref{eq:alpha-Tc}, we can obtain $\alpha(T=T_\mathrm{c}^\ast)\approx$ \qty{68.89e3}{\W\m^{-2}\kelvin^{-1}} and $T_\mathrm{c}^\ast\approx$ \qty{7.707}{\K}. Thus, $B=\alpha(T=T_\mathrm{c}^\ast)/(T_\mathrm{c}^\ast)^3\approx$ \qty{102.8}{\W\m^{-2}\kelvin^{-4}}. \\
$c$ is the heat capacity of the NbTiN film, which is defined as
\begin{equation}
    c=c_\mathrm{e}+c_\mathrm{ph}
\end{equation}
where $c_\mathrm{e}$ is the electron heat capacity and $c_\mathrm{ph}$ is the phonon heat capacity. They are calculated using the following expressions:
\begin{equation}
    c_\mathrm{e}=
    \begin{cases}
    \gamma T,& \text{in normal state,}\\
    A\mathrm{e}^{-\Delta(T)/\mathrm{k}_\mathrm{B}T},& \text{in superconducting state}.
    \end{cases}
\end{equation}
and
\begin{equation}
    c_\mathrm{ph}=c_\mathrm{ph0}T^3.
\end{equation}
The superconducting energy gap is estimated with the following expression\cite{Gross_SCgap} 
\begin{equation}
    \Delta(T))=1.76\mathrm{k}_\mathrm{B}T_\mathrm{c}\tanh(\frac{\pi}{1.76}\sqrt{\frac{2}{3}\times1.43(\frac{T_\mathrm{c}}{T}-1)}),
\end{equation}
where $\mathrm{k}_\mathrm{B}\approx$\qty{1.38e-23}{\joule\per\K} is the Boltzmann constant. $\gamma\approx$\qty{151.5}{\joule\per\cubic\meter\square\kelvin} and $c_\mathrm{ph0}\approx$\qty{12.849}{\joule\per\cubic\meter\kelvin\tothe{4}} are estimated from the reported data of the 9-nm NbTiN film in \citeauthor{Sidorova_2022}.\cite{Sidorova_2022} According to the literature, $A=2.43\gamma T_\mathrm{c}$.\cite{tinkham2004introduction} \\

\section{Efficiency recovery measurement}
\begin{figure}[htbp]
    \centering
    \includegraphics[width =0.75\textwidth]{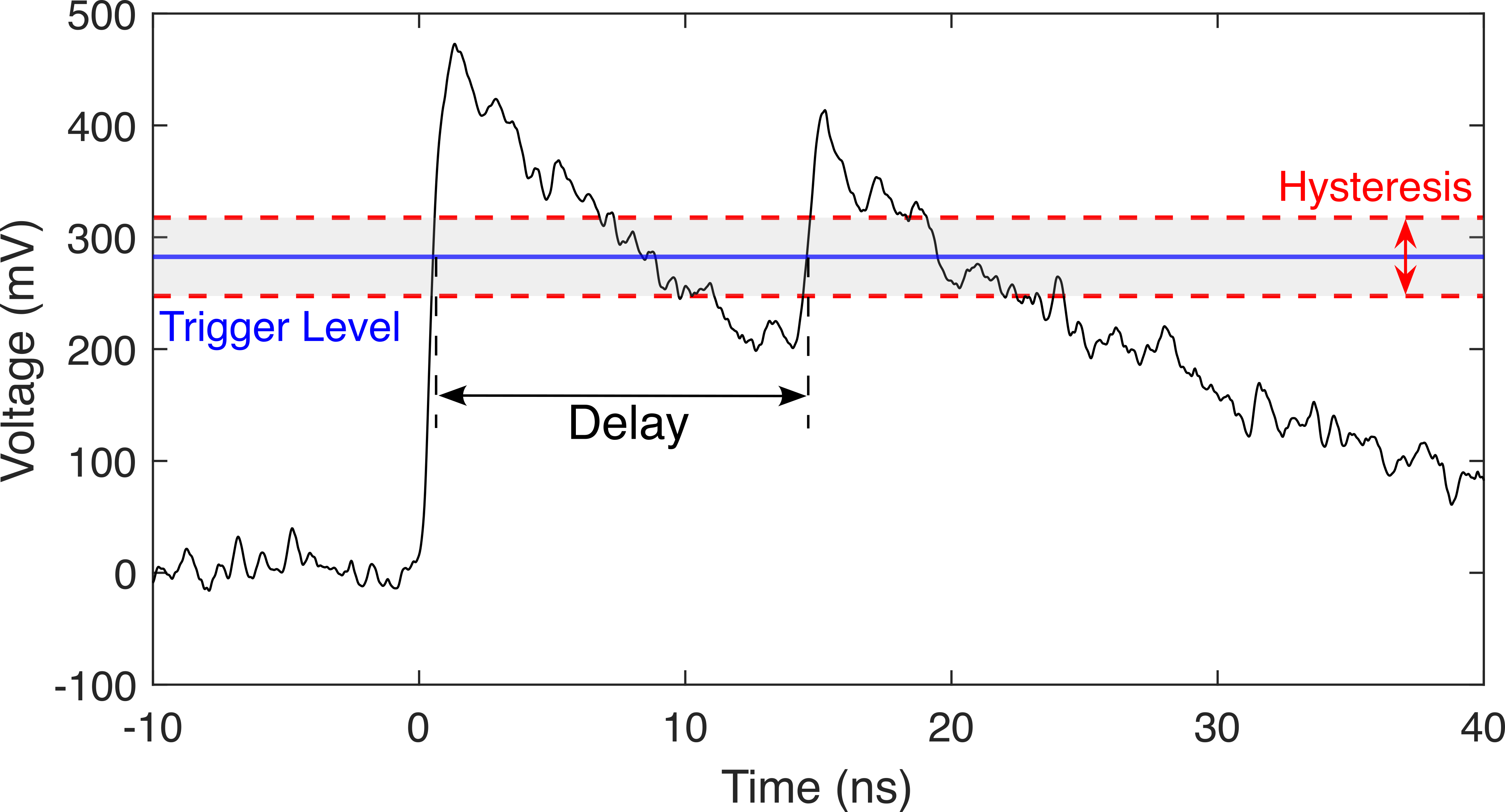}
    \caption{Schematic of the delay time measurement on the output waveform of two successive detection events.}

    \label{fig:meas_waveform}
\end{figure}

In this work, we experimentally measure the efficiency recovery using continuous laser sources. Following the Poissonian statistics of photon distribution from laser sources, the probability of a subsequent photon incidence is independent of the time interval between two photons. Therefore, the distribution of the delay time between two detection events reflects the time-dependent recovery of the detection efficiency after one detection event. As Fig. \ref{fig:meas_waveform} shows, rising edges across certain trigger level are recognized by an oscilloscope (Teledyne LeCroy WaveRunner 640Zi). The time difference between two successive rising slopes at the trigger level is measured as the delay time. Since the second pulse detected after a shorter delay has a lower amplitude due to the partially recovered bias current, the trigger level is empirically set to \qty{60}{\percent} of the first pulse amplitude. To rule out errors induced by electrical noises or fluctuations, we set a hysteresis band of \qty{\pm 35}{mV} around the trigger level. Over 200,000 measurements are acquired and stored as a histogram (as Fig. 3(b) shows) by the oscilloscope.

\section{\label{HCR} High-count-rate Measurements}
\begin{figure*}[!htbp]
    \centering
    \includegraphics[width =1\textwidth]{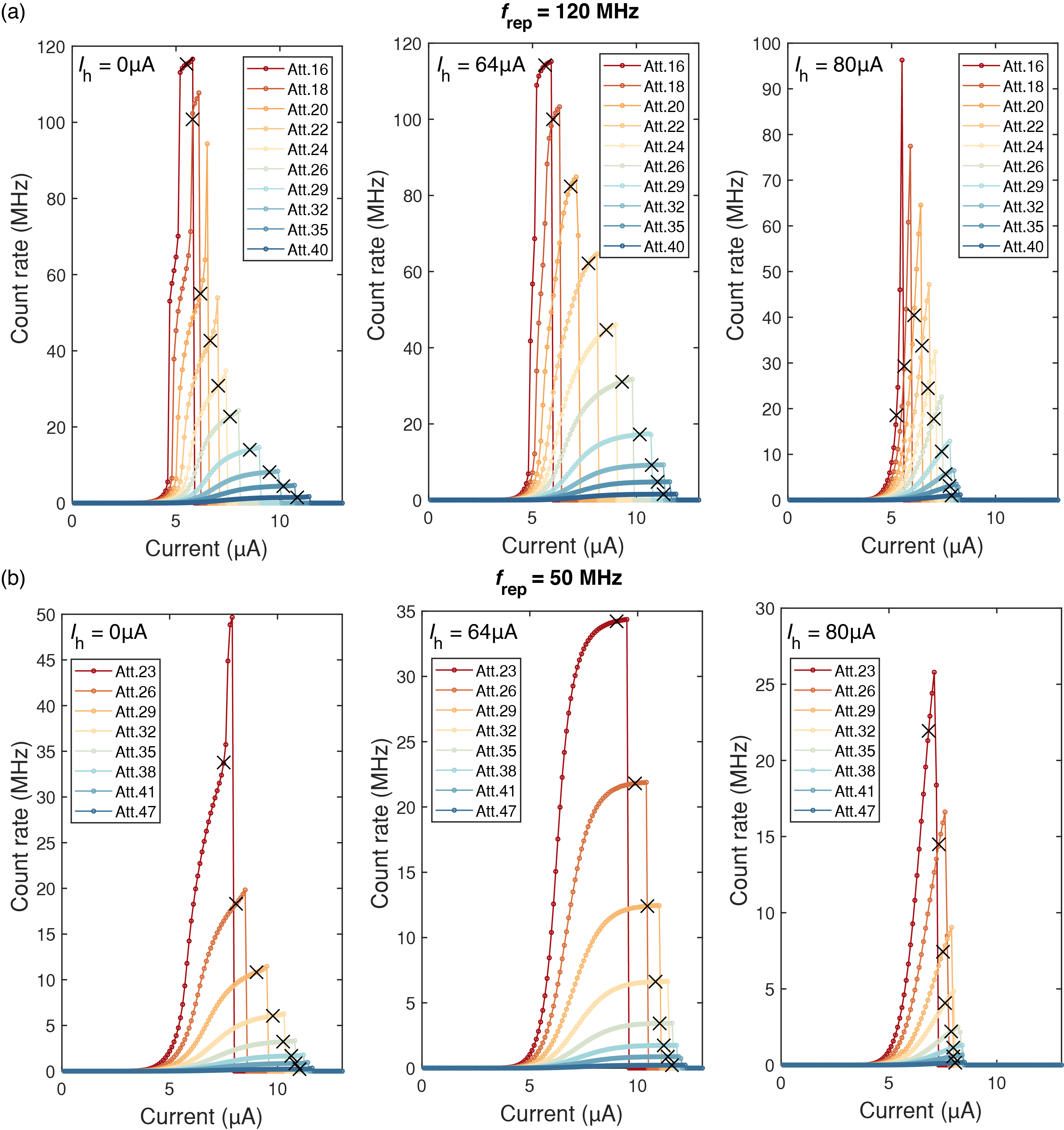}
    \caption{Relationship between bias currents and count rates at various attenuations of light and heater currents $I_\mathrm{h}$ with a repetition rate of (a) $f_\mathrm{rep}=$ \qty{120}{\MHz} and (b) $f_\mathrm{rep}=$ \qty{50}{\MHz}. Black crosses indicate the count rates at \qty{95}{\%} of the switching current, which are used to calculate detection efficiencies in Fig. \ref{fig:HCR_etta}.}
    \label{fig:HCR_IC}
\end{figure*}

\begin{figure*}[!htbp]
    \centering
    \includegraphics[width =1\textwidth]{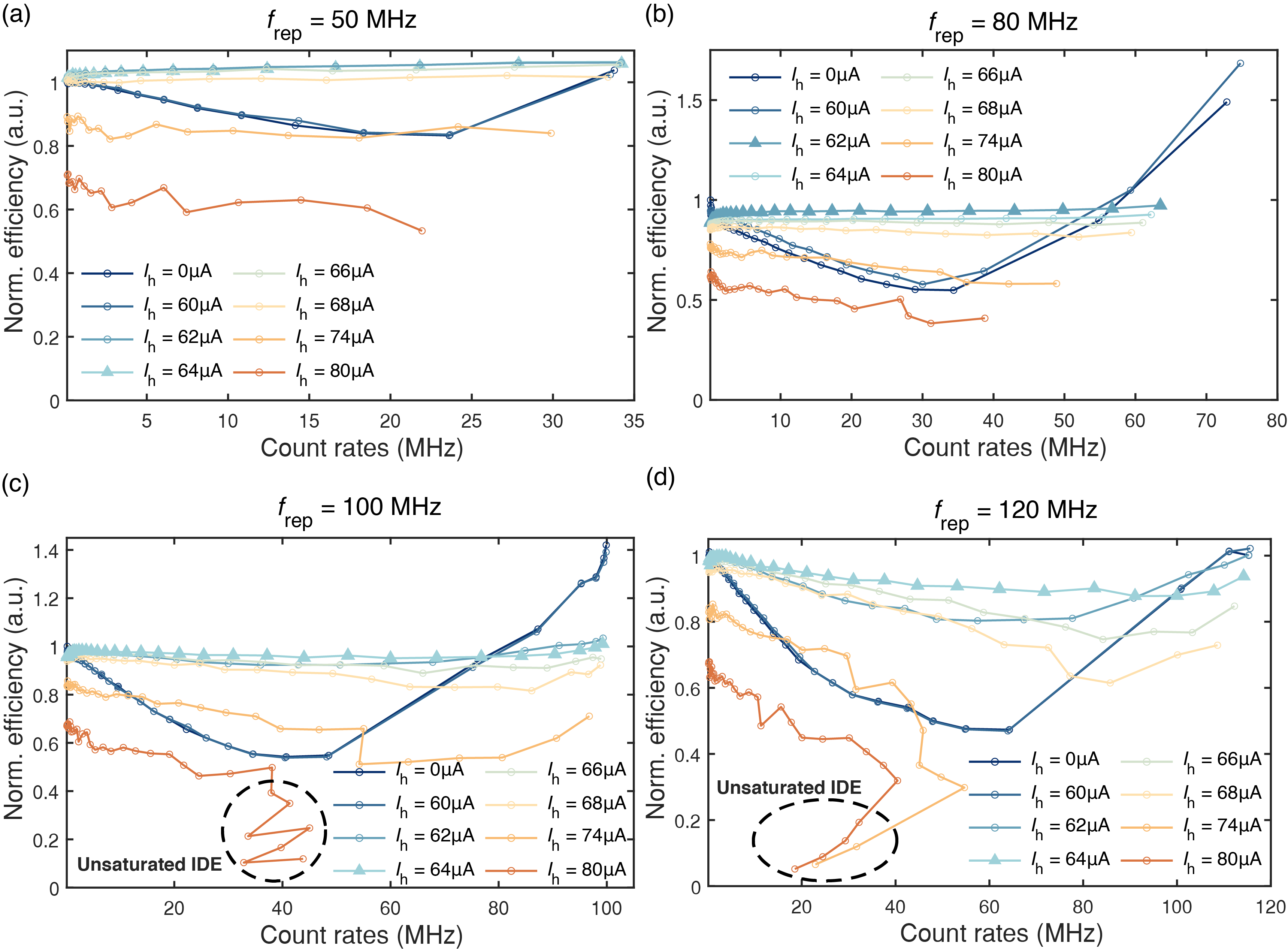}
    \caption{Normalized detection efficiency as a function of detection count rates at various heater currents $I_\mathrm{h}$ with a repetition rate of (a) $f_\mathrm{rep}=$ \qty{50}{\MHz}, (b) $f_\mathrm{rep}=$ \qty{80}{\MHz}, (c) $f_\mathrm{rep}=$ \qty{100}{\MHz}, and (d) $f_\mathrm{rep}=$ \qty{120}{\MHz}. Measurements at the optimal heater currents are marked as triangles for each repetition rate. Data within the dashed circles is caused by the unsaturated internal detection efficiency (IDE).}
    \label{fig:HCR_etta}
\end{figure*}

In high-count-rate measurements, the incident photon rate could exceed a hundred megahertz, which is higher than $1/\tau_\mathrm{out}$. Due to the lack of photon-number-resolving capability, comparing the measured photon count rate with the incident photon rate under continuous light illumination can result in an overestimation or underestimation of the detection efficiency. Therefore, we employ the following method to evaluate the detection efficiency at high count rates using a pulsed laser. \\  
\indent The laser pulses are generated by modulating a laser diode output with an arbitrary waveform generator (AWG). The pulse duration is set to be around \qty{3.3}{\nano\second}, which is limited by the AWG settings. Since it is substantially shorter than the recovery time, only one event can be triggered per optical pulse despite potential multi-photon absorption. Rather than counting the amount of detected photons, we determine the detection efficiency of the SNSPD $\eta$ from the possibility that zero photon is detected in one optical pulse.\\
According to the Poisson distribution, the possibility to have $n$ photons in a pulse can be expressed as 
\begin{equation}
    \label{eq:Poisson}
    P(n)=\frac{\mu^n}{n!}e^{-\mu},
\end{equation}
where $\mu$ is the average number of photons in a pulse. The average number of photons is calculated using
\begin{equation}
\mu=\frac{\lambda P_{\mathrm{in}}}{hcf_{\mathrm{rep}}},   
\end{equation}
where $\lambda$ is the laser wavelength, $P_{\mathrm{in}}$ is the input laser power, $h\approx$ \qty{6.62607e-34}{\joule\per\hertz} is the Planck constant, and $c\approx$ \qty{2.99792e9}{\meter\per\second} is the speed of light. Firstly, the possibility where the number of detection events $d$ for an optical pulse equals $0$ can be described analytically as
\begin{eqnarray}
\label{eq:P_d=0 math}
    P(d=0)&&=P(n=0)+P(n\neq 0)(1-\eta)^n\nonumber\\
    &&=\sum_{n=0}^{\inf}P(n)(1-\eta)^n\nonumber\\
    &&=\sum_{n=0}^{\inf}\frac{\mu^n}{n!}e^{-\mu}(1-\eta)^n\nonumber\\
    &&=e^{-\eta\mu}.
\end{eqnarray}
Second, $P(d=0)$ can also be obtained from the experimental result, which is written as
\begin{equation}
\label{eq:P_d=0 exp}
P(d=0)=1-\frac{\mathrm{PCR}}{f_{\mathrm{rep}}},
\end{equation}
where $\mathrm{PCR}$ is the photon count rate measured by the SNSPD and $f_{\mathrm{rep}}$ is the repetition rate of the pulsed laser. Combining Eq. \ref{eq:P_d=0 exp} and Eq. \ref{eq:P_d=0 math}, we can derive that
\begin{equation}
    \eta=-\frac{\ln(1-\frac{\mathrm{PCR}}{f_{\mathrm{rep}}})}{\mu}.
\end{equation}
\indent By applying various attenuation levels on the input laser source, we measure sets of count-rate curves versus bias currents, as shown in Fig. \ref{fig:HCR_IC}. Count rates at \qty{95}{\%} of the switching currents are used to derive the relationship between detection efficiencies and count rates in Fig. 4 and \ref{fig:HCR_etta} at different repetition rates. Since the selected count rate might fall on the unsaturated region due to the decline of switching current, the calculated efficiency can fluctuate or decrease as the region highlighted by dashed circles for $I_\mathrm{h}=$ \qty{80}{\uA} and $I_\mathrm{h}=$ \qty{120}{\uA} in Fig. \ref{fig:HCR_etta} (c) and (d) shows. Similar to the analysis of Fig. 4, the detection efficiency at high count rates is substantially improved at $I_\mathrm{h}$ around \qty{64}{uA}, which is consistent with the result of efficiency recovery measurements. The measured optimal heater current slightly shifts for $f_\mathrm{rep}=$ \qty{80}{MHz}, which can be caused by incident power drift during experiments due to laser power fluctuations or fiber instability. The normalized detection efficiency exceeding unity at high count rates can be due to multiple detection events within one laser pulse or oscillations caused by the parasitic discharging current of the readout capacitor.

\clearpage

\bibliography{bibl}